\documentclass[10pt,conference,compsocconf]{IEEEtran}
\usepackage{latexsym,amsmath,amsfonts,graphics}
\ifCLASSINFOpdf
\else
\fi
\begin{document}
\def\QEDclosed{\mbox{\rule[0pt]{1.4ex}{1.4ex}}}
\def\QED{\QEDclosed}
\title{Topological Properties of an Exponential Random Geometric Graph Process$^1$}

\author{\IEEEauthorblockN{Yilun Shang}
\IEEEauthorblockA{Department of Mathematics\\
Shanghai Jiao Tong University\\
Shanghai 200240, China\\
Email: shyl@sjtu.edu.cn}}
\maketitle

\begin{abstract}
In this paper, we consider a one-dimensional random geometric graph
process with the inter-nodal gaps evolving according to an
exponential AR(1) process, which may serve as a mobile wireless
network model. The transition probability matrix and stationary
distribution are derived for the Markov chains in terms of network
connectivity and the number of components. We characterize an
algorithm for the hitting time regarding disconnectivity. In
addition, we also study topological properties for static snapshots.
We obtain the degree distributions as well as asymptotic precise
bounds and strong law of large numbers for connectivity threshold
distance and the largest nearest neighbor distance amongst others.
Both closed form results and limit theorems are
provided.\footnotetext[1]{This paper is an extended version of a
conference paper ``Exponential Random Geometric Graph Process Models
for Mobile Wireless Networks'' pp. 56--61, presented in
International Conference on Cyber-Enabled Distributed Computing and
Knowledge Discovery, 2009.}
\end{abstract}

\begin{IEEEkeywords}
random geometric graph; autoregressive process; component;
connectivity; mobile network.
\end{IEEEkeywords}

\IEEEpeerreviewmaketitle

\section{Introduction}

Many randomly deployed networks, such as wireless sensor networks,
are properly characterized by random geometric graphs (RGGs). Given
a specified norm on the space under consideration, an RGG is usually
obtained by placing a set of $n$ vertices independently at random
according to some spatial probability distribution and connecting
two vertices by an edge if and only if their distance is less than a
critical cutoff $r$. Topological properties of RGGs are
comprehensively summarized in \cite{15}; also see \cite{20} for a
recent survey in the context of wireless networks. Although
extensive simulations and empirical studies are performed in
dynamical RGGs, analytical treatments of topological properties are
merely done in static RGGs in the previous work. A recent paper
\cite{13} is a remarkable exception, in which the authors conduct
the first analytical research on the connectivity of mobile RGG in
the torus $[0,1)^2$. In this paper, we will also present analytical
results and consider a one-dimensional exponential RGG process
$G(t,r,\Lambda)$ evolving with time, where vertices are randomly
placed along a semi-infinite line. One-dimensional exponential RGGs
have been recently investigated by some authors
\cite{10}--\cite{11}, which offer a significant variant from the
familiar uniformly $U[0,1]$ distributed nodes, see e.g. \cite{20},
\cite{21}--\cite{50} and references therein.

In \cite{14}, the distributions of distances between successive
vertices rather than those of vertices themselves are examined, and
as it is stated in the same paper, this assumption is more natural
since ``\textit{sensors are usually thrown one by one along a
trajectory of a vehicle}.'' We will then follow suit, and assume
exponential distributions for inter-nodal distances of the graph
process $G(t,r,\Lambda)$. Every segment between two successive
vertices is supposed to evolve following a stationary TEAR(1)
process \cite{16} with exponential marginal. This linear process has
no zero-defect and thus surpasses the elementary AR(1) process
involved in \cite{11}. We believe such a mobile scheme has broad
potential applications due to the flexible double randomness
mechanism (see Section II). Since the evolution of connectivity and
the number of components in $G(t,r,\Lambda)$ are both Markovian, we
will address the transition probabilities and limiting distributions
of these two processes $G_t$ and $G'_t$ by employing Markov chain
theory \cite{9,8}. It is worth noting that there are several Markov
chains coupled in our model stemming from the first order
autoregressive properties endowed in the evolution of inter-nodal
distances.

In addition to dynamical properties, we also establish static
properties for fixed $t$. Vertices in $G(t,r,\Lambda)$, for any
given $t$, form nearly a Poisson point process (more precisely, a
continuous time pure birth Markov process). The connectivity of a
Poisson RGG is well-studied in the literature (see e.g.
\cite{3}--\cite{4}), especially in the context of ad hoc networks. We
will investigate some topological properties basically along the
lines of \cite{10}. In our opinion, the aforementioned simple idea
in \cite{14} reflects a conception of one step ``memory''
essentially. We show (in Theorem 4) that ``1-step memory'' +
``growth'' are not enough to produce power law distribution
reminiscent of the architecture of Polya urn process, where
typically infinite memory generates the power law \cite{1}.

Both finite and asymptotic analysis are given in this paper. We
remark here that exact solutions are important since the asymptotic
results can not be applied to real networks when not knowing the
rate of convergence.

The rest of this paper is organized as follows. Section II provides
the definition of the exponential RGG process and some
preliminaries. Section III deals with the evolutionary properties of
$G(t,r,\Lambda)$, including the transition probability, the
stationary distribution and the hitting time for disconnectivity. In
Section IV, we present static topological properties of
$G(t,r,\Lambda)$ for fixed $t$. The degree distribution and strong
laws of connectivity and the largest nearest neighbor distances are
given among other things. In Section V, some concluding remarks and
future research topics are discussed.

\section{Model and Preliminaries}

The RGG process $G(t,r,\Lambda)$ is constructed as a discrete time
process with $n$ vertices deployed in one dimension on $[0,\infty)$.
Let $X_1^t,\cdots,X_n^t$ denote the vertices of the network at time
$t$, for $t\ge0$. Set $Y_l^t:=X_{l+1}^t-X_l^t$, for
$l=1,2,\cdots,n-1$ and $Y_0^t:=X_1^t$; see Fig.1 for an
illustration. We may envision time evolving upward along the $t$-axis and $n$
vertices possibly growing along the $x$-axis.

For $0\le p<1$, we assume that $\{Y_l^t\}$ evolves following:
\begin{equation}
Y_l^{t+1}=
\begin{cases}Y_l^t+\varepsilon_l^t & w.p.\quad p\\ \varepsilon_l^t & w.p.
\quad 1-p\label{1}
\end{cases}
\end{equation}
where the innovation sequences $\{\varepsilon_l^t\}_{t\ge0}$ consist
of i.i.d. nonnegative random variables. The behavior of this
autoregressive process $\{Y_l^t\}_{t\ge0}$ is characterized by runs
of rising values (with geometrically distributed run length) when
choosing $Y_l^t+\varepsilon_l^t$, followed by a sharp fall when
choosing $\varepsilon_l^t$ without inclusion of the previous values.
Furthermore, we assume that $Y_l^t$, $l=0,1,\cdots,n-1$ are
independent for any t.

In particular, we set $\varepsilon_l^t:=(1-p)Z_l^t$, where
$Z_l^t\sim Exp(\lambda_l)$ is an exponential random variable with
mean $\lambda_l^{-1}>0$. Let
$\Lambda:=\{\lambda_0,\lambda_1,\cdots,\lambda_{n-1}\}$. In this
case, as is shown in \cite{16}, the above TEAR(1) process
$\{Y_l^t\}_{t>0}$ would be a stationary sequence of marginally
exponentially distributed random variables with parameter
$\lambda_l$, assuming that the initial inter-nodal gaps $Y_l^0$ are
exponentially distributed with parameter $\lambda_l$. That means
$Y_l^t\sim Exp(\lambda_l)$. In this case, the auto correlation
function of $\{Y_l^t\}$ is $\mathrm{Corr}(Y_l^t,Y_l^{t+j})=p^j$,
being nonnegative. Reference \cite{17} showed that (\ref{1}) is
stationary for each $0\le p<1$ iff $Y_l^t$ is geometrically
infinitely divisible. For further extension and discussion of
(\ref{1}) we refer the reader to \cite{18}.

\textit{Remark 1:} Vertices in snapshot of $G(t,r,\Lambda)$ yield a
counting process with inter-nodal distances having distribution
$Exp(\lambda_l)$, while in standard exponential RGG, the
corresponding distributions are relevant to $n$ the total number of
vertices (see Lemma 1 in \cite{10}) relying on the global
information.

\textit{Remark 2:} Notice that the cutoff $r=r(n,t)$ may depend on
$n$ and $t$. However, we restrict ourselves to fixed $r$ in order to
keep calculations clear though some results may be generalized
without much effort. The popular assumption
$\lim_{n\rightarrow\infty}r(n)=0$ is not necessary here in virtue of
unbounded support resulting from inter-nodal spacing.

\begin{figure}[!t]
\centering
\scalebox{0.8}{\includegraphics[5cm,24.5cm][30cm,26.5cm]{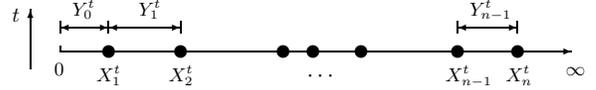}}
\caption{One-dimensional exponential RGG process model}
\label{fig_sim}
\end{figure}

\section{Evolutionary Properties of
$G(t,r,\Lambda)$}

\subsection{Stationary Distribution of $G_t$}

Let us denote by $\mathcal{C}_t$ and $\mathcal{D}_t$ the events that
$G(t,r,\Lambda)$ is connected and disconnected at time $t$,
respectively. Define $G_t$ as a discrete time stochastic process
describing connectivity of the graph process $G(t,r,\Lambda)$.
Therefore $\mathcal{C}_t=\{G_t=``\mathrm{conneted}$''$\}$ and
$\mathcal{D}_t=\{G_t=``\mathrm{disconneted}$''$\}$. It is easy to
see that $G_t$ is a homogeneous Markov chain, assuming the cutoff
$r$ is independent of $t$. We abbreviate as usual the states as
1=``connected'' $(\mathcal{C})$ and 2=``disconnected''
$(\mathcal{D})$. Our main results in this section then read as
follows:

\textit{Theorem 1:} $G_t$ is a time-reversible, homogeneous finite
Markov chain, with one step transition probability matrix
$$
P(n)=\left(\begin{array}{cc} p_{11} & p_{12}\\
p_{21} & p_{22}
\end{array}\right),
$$
where
\begin{equation}
p_{11}=\prod_{l=1}^{n-1}\bigg(1-\frac{(1-p)e^{-\lambda_lr}\big(1-e^{-\frac{\lambda_lr}{1-p}}\big)}{1-e^{-\lambda_lr}}\bigg)\label{2},
\end{equation}
\begin{eqnarray}
p_{21}&=&\frac{\sum\limits_{\emptyset\not=A\subseteq[n-1]}(1-p)\prod\limits_{l\in
A}e^{-\lambda_lr}\big(1-e^{-\frac{\lambda_lr}{1-p}}\big)}{1-\prod\limits_{l=1}^{n-1}(1-e^{-\lambda_lr})}\nonumber\\
& &\cdot\hspace{-2mm}\prod\limits_{l\in [n-1]\backslash
A}\hspace{-3mm}\big(1-e^{-\lambda_lr}-(1-p)e^{-\lambda_lr}\big(1-e^{-\frac{\lambda_lr}{1-p}}\big)\big),\nonumber\\
\hfill\label{3}
\end{eqnarray}
$p_{12}=1-p_{11}$ and $p_{22}=1-p_{21}$.

\textit{Proposition 1:} $G_t$ has a unique stationary distribution
$\pi(n)=(\pi_1(n),\pi_2(n))$, where
\begin{equation}
\begin{split}
\pi_1(n)&=\frac{(1-p_{22})^2}{p_{11}(1-p_{22})^2+p_{21}p_{12}(2-p_{22})}\\\mathrm{and}\quad
\pi_2(n)&=\frac{(1-p_{11})^2}{p_{22}(1-p_{11})^2+p_{12}p_{21}(2-p_{11})}.\label{4}
\end{split}
\end{equation}

\textit{Proposition 2:} Suppose $\lambda_l\equiv\lambda$, for
$l=0,1,\cdots,n-1$. Let $P(\infty)$ be the transition probability
matrix of $G_t$ as $n$ tends to infinity, and $\pi(\infty)$ the
(unique) stationary distribution corresponding to $P(\infty)$. Then
$\pi(\infty)=(0,1)$ and
$$
\lim_{n\rightarrow\infty}\pi(n)P(n)=\pi(\infty)P(\infty).
$$

Proposition 2 implies that we can swap the order of obtaining
stationary distribution and taking limit w.r.t. $n$.

\textit{Proof of Theorem 1:} The probability density function of
$\varepsilon_l^t$ can be shown to be given by
$f_l(s)=\frac{\lambda_l}{1-p}e^{-\lambda_ls/(1-p)}1_{[s>0]}$. Also,
the conditional density function for $Y_l^t$ in the connected
network is
$g_{Y_l|\mathcal{C}}(y)=\frac{\lambda_le^{-\lambda_ly}}{1-e^{-\lambda_lr}}1_{[0<y<r]}$,
since the connectivity of network means $Y_l^t<r$ for all $l$. By
independence property, we have
$p_{11}=P(\mathcal{C}_{t+1}|\mathcal{C}_t)=\prod_{l=1}^{n-1}P(Y_l^{t+1}<r|Y_l^t<r)$.
Our aim now turns to evaluate the probability
$P(Y_l^{t+1}<r|Y_l^t<r)$. Let $V_l^t\sim Ber(p)$ independently, then
the scheme (\ref{1}) becomes
\begin{equation}
Y_l^{t+1}=\varepsilon_l^t+V_l^tY_l^t.\label{5}
\end{equation}
Let $\widetilde{Y}_l^{t+1}$ denote $Y_l^{t+1}$ conditional on
$\{Y_l^t<r\}$. For a nonnegative random variable $X$ with density
function $f(x)$, the Laplace-Stieltjes transform is defined by
$\mathcal{L}(X)(s)=\mathcal{L}(f)(s)=\int_0^{\infty}f(x)e^{-sx}\mathrm{d}x$.
We have by (\ref{5}),
\begin{eqnarray*}
\mathcal{L}(\widetilde{Y}_l^{t+1})(s)&=&\mathcal{L}(\varepsilon_l^t)(s)\cdot\mathcal{L}(V_l^t\widetilde{Y}_l^t)(s)\\
&=&\int_0^{\infty}e^{-su}\frac{\lambda_l}{1-p}e^{-\frac{\lambda_lu}{1-p}}\mathrm{d}u\\
& &\cdot\int_0^re^{-sy}\Big((1-p)\delta(y)+\frac{p\lambda_le^{-\lambda_ly}}{1-e^{-\lambda_lr}}\Big)\mathrm{d}y\\
&=&\frac{\lambda_l}{\lambda_l+s(1-p)}\\
&
&\cdot\Big((1-p)+\frac{p\lambda_l(1-e^{-(\lambda_l+s)r})}{(s+\lambda_l)(1-e^{-\lambda_lr})}\Big)
\end{eqnarray*}
where $\delta(y)$ is the Dirac-delta function. Inverting the above
to get
\begin{eqnarray*}
\mathcal{L}^{-1}(\mathcal{L}(\widetilde{Y}_l^{t+1}))(y)&=&\lambda_le^{-\frac{\lambda_ly}{1-p}}1_{[y>0]}+\frac{2\lambda_le^{-\frac{\lambda_l(2-p)y}{2(1-p)}}}{1-e^{-\lambda_lr}}\\
& &\cdot\mathrm{sinh}\Big(\frac{\lambda_lpy}{2(1-p)}\Big)1_{[y>0]}\\
&
&-\frac{2\lambda_le^{-\lambda_l\big(r+\frac{(2-p)(y-r)}{2(1-p)}\big)}}{1-e^{-\lambda_lr}}\\
&
&\cdot\mathrm{sinh}\Big(\frac{\lambda_lp(y-r)}{2(1-p)}\Big)1_{[y>r]}.
\end{eqnarray*}
Hence
\begin{eqnarray}
P(Y_l^{t+1}<r|Y_l^t<r)&=&\int_0^r\mathcal{L}^{-1}(\mathcal{L}(\widetilde{Y}_l^{t+1}))(y)\mathrm{d}y\nonumber\\
&=&1-\frac{(1-p)e^{-\lambda_lr}\big(1-e^{-\frac{\lambda_lr}{1-p}}\big)}{1-e^{-\lambda_lr}}\nonumber\\
\hfill\label{6}
\end{eqnarray}
which gives (\ref{2}).

Let $\emptyset\not=A\subseteq[n-1]:=\{1,2,\cdots,n-1\}$. Denote the
event $E_A:=\{Y_l^t>r,\forall l\in A;Y_l^t<r,\forall
l\in[n-1]\backslash A\}$, then we have
\begin{eqnarray*}
P(\mathcal{C}_{t+1}|E_A)&=&\hspace{-2pt}\prod_{l\in
A}P(Y_l^{t+1}<r|Y_l^t>r)\\
& &\cdot\prod_{l\in [n-1]\backslash
A}P(Y_l^{t+1}<r|Y_l^t<r)\\
&=&\hspace{-2pt}\prod_{l\in
A}(1-p)\big(1-e^{-\frac{\lambda_lr}{1-p}}\big)\\
& &\hspace{-2pt}\cdot\hspace{-5pt}\prod_{l\in[n-1]\backslash
A}\hspace{-3mm}\Big(1-\frac{(1-p)e^{-\lambda_lr}\big(1-e^{-\frac{\lambda_lr}{1-p}}\big)}{1-e^{-\lambda_lr}}\Big).
\end{eqnarray*}
Here we used the expression
$P(Y_l^{t+1}<r|Y_l^t>r)=(1-p)\big(1-e^{-\frac{\lambda_lr}{1-p}}\big)$.
Since $P(E_A)=\prod_{l\in
A}e^{-\lambda_lr}\prod_{l\in[n-1]\backslash A}(1-e^{-\lambda_lr})$
and $P(\mathcal{D}_t)=1-\prod_{l=1}^{n-1}(1-e^{-\lambda_lr})$,
(\ref{3}) follows by noting that
$$
p_{21}=P(\mathcal{C}_{t+1}|\mathcal{D}_t)=\sum_{\emptyset\not=A\subseteq[n-1]}P(\mathcal{C}_{t+1}|E_A)\cdot
P(E_A)/P(\mathcal{D}_t).
$$
$G_t$ is time-reversible by standard results of Markov chains
\cite{9}. \hspace*{\fill}~\QED

\textit{Proof of Proposition 1:} Since $G_t$ is an irreducible
finite Markov chain, $\mathcal{C}$ and $\mathcal{D}$ are both
positive recurrent. Also since they are both non-periodical,
$\mathcal{C}$ and $\mathcal{D}$ are ergodic states. Set
$T_{ij}:=\min\{k:k\ge1,G_k=j,G_0=i\}$, for $i,j\in\{1,2\}$. If the
righthand side of the above definition is $\emptyset$, set
$T_{ij}=\infty$. The first hitting probability is then given by
$f_{ij}^{(k)}=P(T_{ij}=k|G_0=i)$.

By a standard result from \cite{8}, an irreducible ergodic Markov
chain has unique stationary distribution $\pi(n)$, and $\pi_i(n)$ is
given by $\pi_i(n)=1/\sum_{k=1}^{\infty}kf_{ii}^{(k)}$, for $i=1,2$
in the present case. Thereby, (\ref{4}) follows easily from the
facts $f_{11}^{(1)}=p_{11}$,
$f_{11}^{(k)}=p_{21}p_{22}^{k-2}p_{12}$, for $k\ge2$; and
$f_{22}^{(1)}=p_{22}$, $f_{22}^{(k)}=p_{12}p_{11}^{k-2}p_{21}$, for
$k\ge2$. \hspace*{\fill}~\QED

\textit{Proof of Proposition 2:} When $\lambda_l\equiv \lambda$, the
righthand side of expression (\ref{6}) belongs to interval $(0,1)$.
Hence $p_{11}$ tends to 0 as $n\rightarrow\infty$ in view of
(\ref{2}). Since $(1-p)e^{-\lambda r}\big(1-e^{-\frac{\lambda
r}{1-p}}\big)+\big(1-e^{-\lambda r}-(1-p)e^{-\lambda
r}\big(1-e^{-\frac{\lambda r}{1-p}}\big)\big)=1-e^{-\lambda r}<1$,
$p_{21}$ tends to 0 as $n\rightarrow\infty$ by the binomial theorem
and (\ref{3}). Then we have
$P(\infty)=\left(\begin{array}{cc} 0 & 1\\
0 & 1
\end{array}\right).$
In this case, $\mathcal{C}$ is a transient state and $\mathcal{D}$
is an absorbing and positive recurrent state. By a standard result
(see e.g. \cite{8}), the stationary distribution corresponding to
$P(\infty)$ exists and is unique. Direct calculation gives
$\pi(\infty)=(0,1)$. It is straightforward to verify that
$\pi(n)\rightarrow\pi(\infty)$ as $n$ tends to infinity. The theorem
is thus concluded by exploiting the relation $\pi P=\pi$.
\hspace*{\fill}~\QED

\subsection{Transition Probability Matrix of
$G'_t$}

In this section, we show a refinement stochastic process $G'_t$ from
$G_t$. To be precise, let $\{G'_t=i\}$ denote the event that
$G(t,r,\Lambda)$ has $i$ components at time $t$, for $1\le i\le n$.
Therefore, $G'_t$ is a homogeneous Markov chain with state space
$[n]$. It's clear that $\{G'_t=1\}=\mathcal{C}_t$.

Let the transition probabilities of $G'_t$ be
$p'_{ij}:=P(G'_{t+1}=j|G'_t=i)$. Set $A,B\subseteq[n-1]$ with
$|A|=i-1$ and $|B|=j-1$, $1\le i,j\le n$. Denote the event
$E_A:=\{Y_l^t>r,\forall l\in A;Y_l^t<r,\forall l\in[n-1]\backslash
A\}$ and similarly for $E_B$. We obtain by the total probability
formula,
\begin{equation}
p'_{ij}=\sum_{A,B\subseteq[n-1]\atop |A|=i=1,|B|=j-1}P(E_B|E_A)\cdot
P(E_A)/P(G'_t=i),\label{7}
\end{equation}
for $1\le i,j\le n$. We have derived $P(E_A)$ in the proof of
Theorem 1, and
$P(G'_t=i)=\sum\limits_{A\subseteq[n-1],|A|=i-1}P(E_A)$. To evaluate
(\ref{7}), we still need the probability $P(E_B|E_A)$, but it is
also at hand already:
\begin{eqnarray*}
P(E_B|E_A)&=&\prod_{l\in A\cap B}P(Y_l^{t+1}>r|Y_l^t>r)\\
& &\cdot\prod_{l\in
A\backslash B}P(Y_l^{t+1}<r|Y_l^t>r)\\
& &\cdot\prod_{l\in B\backslash
A}P(Y_l^{t+1}>r|Y_l^t<r)\\
& &\cdot\prod_{l\in [n-1]\backslash A\cup B}P(Y_l^{t+1}<r|Y_l^t<r).
\end{eqnarray*}
The second and fourth terms in the above expression have been
obtained in the proof of Theorem 1, and clearly $
P(Y_l^{t+1}>r|Y_l^t>r)=1-P(Y_l^{t+1}<r|Y_l^t>r),
P(Y_l^{t+1}>r|Y_l^t<r)=1-P(Y_l^{t+1}<r|Y_l^t<r). $ Now we arrive at
the main result.

\textit{Theorem 2:} The transition probability matrix of $G'_t$ is
$P'=(p'_{ij})_{n\times n}$, which is given by (\ref{7}).

Of course, we have $p'_{11}=p_{11}$ and
$\sum_{j=2}^np'_{1j}=p_{12}$. Since $G'_t$ is an irreducible ergodic
chain, it has a unique stationary distribution which may be deduced
analogously as in Section III.A.

\subsection{Hitting Time for Disconnectivity}

Suppose $\mathcal{C}_t$ holds at time $t$, and we will consider the
Markov chain $G_t$. Denote $T:=\min\{k:k\ge1,\mathcal{D}_{t+k}\
\mathrm{holds}\}$, then $T$ is the hitting time for disconnectivity.
We may obtain the expectation of $T$ using the transition
probabilities derived in Section III.A by a routine approach
\cite{8}. In this section, we will instead depict an algorithm for
getting the distribution of $T$ directly.

The event $\{T>k\}$ is equivalent to $\{Y_l^{t+1}<r$, $Y_l^{t+2}<r$,
$\cdots$,$Y_l^{t+k}<r$, $\forall$ $1\le l\le n-1\}$. In view of
(\ref{5}), we can interpret the above as follows
\begin{eqnarray*}
Y_l^{t+1}&=&\varepsilon_l^t+V_l^tY_l^t<r,\\
Y_l^{t+2}&=&\varepsilon_l^{t+1}+V_l^{t+1}\varepsilon_l^t+V_l^{t+1}V_l^tY_l^t<r,\\
&\cdots&\\
Y_l^{t+k}&=&\varepsilon_l^{t+k-1}+V_l^{t+k-1}\varepsilon_l^{t+k-2}+\cdots\\
& &+V_l^{t+k-1}\cdots V_l^{t+1}\varepsilon_l^t+V_l^{t+k-1}\cdots
V_l^tY_l^t<r.
\end{eqnarray*}
Set
$U_l^{t+j}:=V_l^{t+j}\varepsilon_l^{t+j-1}+\cdots+V_l^{t+j}\cdots
V_l^{t+1}\varepsilon_l^t+V_l^{t+j}\cdots V_l^tY_l^t$, for $1\le j\le
k-1$ and $U_l^t:=V_l^tY_l^t$. Therefore, conditioned on
$Y_l^t,V_l^t,\cdots,V_l^{t+k-1}$, the probability that the above $k$
inequalities holds simultaneously is shown to be given by
\begin{eqnarray}
\lefteqn{P_l^k(Y_l^t,\{V_l^t,\cdots,V_l^{t+k-1}\})=}\nonumber\\
& \int_0^{r-U_l^t}f_l(\varepsilon_l^t)\mathrm{d}\varepsilon_l^t
\cdots\cdot\int_0^{r-U_l^{t+k-1}}f_l(\varepsilon_l^{t+k-1})\mathrm{d}\varepsilon_l^{t+k-1},&\label{8}
\end{eqnarray}
where $f_l(\cdot)$ is given in the proof of Theorem 1. Denote the
last $i+1$ integrals of (\ref{8}) by $I_{l,k-i}$, $0\le i\le k-1$.
For $i=0$,
$$
I_{l,k}=\int_0^{r-U_l^{t+k-1}}\frac{\lambda_l}{1-p}e^{-\frac{\lambda_ls}{1-p}}\mathrm{d}s=1-e^{-\frac{\lambda_l\left(r-U_l^{t+k-1}\right)}{1-p}}.
$$
For $i=1$,
\begin{eqnarray*}
I_{l,k-1}&=&\int_0^{r-U_l^{t+k-2}}\frac{\lambda_l}{1-p}e^{-\frac{\lambda_l\varepsilon_l^{t+k-2}}{1-p}}
I_{l,k}\mathrm{d}\varepsilon_l^{t+k-2}\\
 &=&1-e^{-\frac{\lambda_l\left(r-U_l^{t+k-2}\right)}{1-p}}\\
 & &-\frac{\lambda_l(r-U_l^{t+k-2})}{1-p}e^{-\frac{\lambda_l\left(r-U_l^{t+k-2}\right)}{1-p}}1_{[V_l^{t+k-1}=1]}\\
&
&-\Big(1-e^{-\frac{\lambda_l\left(r-U_l^{t+k-2}\right)}{1-p}}\Big)e^{-\frac{\lambda_lr}{1-p}}1_{[V_l^{t+k-1}=0]}.
\end{eqnarray*}
In general, for $0\le i\le k-1$,
$$
I_{l,k-i}=\int_0^{r-U_l^{t+k-i-1}}\hspace{-2mm}\frac{\lambda_l}{1-p}e^{-\frac{\lambda_l\varepsilon_l^{t+k-i-1}}{1-p}}I_{l,k-i+1}\mathrm{d}\varepsilon_l^{t+k-i-1}.
$$
We can proceed using this recursive formula by induction and
integration by parts. Notice that
$P_l^k(Y_l^t,\{V_l^t,\cdots,V_l^{t+k-1}\})=I_{l,1}$ from (\ref{8}).

Consequently, given $Y_l^t<r$, the probability that $Y_l^{t+1}<r$,
$Y_l^{t+2}<r$, $\cdots$,$Y_l^{t+k}<r$ all are simultaneously true is
seen to be given by
\begin{eqnarray*}
\widetilde{P}_l^k&:=&\frac{\lambda_l}{1-e^{-\lambda_lr}}\cdot\sum_{i=0}^kp^i(1-p)^{k-i}\\
& &\cdot\sum_{k-\mathrm{vector}\ \xi \atop \mathrm{consisting}\
\mathrm{of}\ k\ 1's, k-i\
0's}\int_0^rP_l^k(y,\xi)e^{-\lambda_ly}\mathrm{d}y.
\end{eqnarray*}
Now we state our result as follows, whose proof is straightforward
at this stage.

\textit{Theorem 3:} Suppose the hitting time $T$ of $G_t$ is defined
as above, then the distribution $P(T\le
k)=1-\prod_{l=1}^{n-1}\widetilde{P}_l^k$ and it's expectation
$ET=\sum_{k=0}^{\infty}\prod_{l=1}^{n-1}\widetilde{P}_l^k$.

In principle, by the truncation of $k$, we may approximate $ET$
arbitrarily close.

\section{Snapshots of $G(t,r,\Lambda)$}

For fixed $t$, we denote by $G(r,\Lambda)$ the static case which can
be regarded as a snapshot of the dynamical process $G(t,r,\Lambda)$.
Also, we omit the superscript $t$ typically, e.g. $Y_l$, etc.

\subsection{Cluster Structure}

Let $P_n(\mathcal{C})$ denote the probability that $G(r,\Lambda)$ is
connected. We have the following result regarding connectivity. The
proof is easy and hence omitted.

\textit{Proposition 3:} We have
$$P_n(\mathcal{C})=\prod_{l=1}^{n-1}(1-e^{-\lambda_lr}).$$ Moreover, suppose there
exists $M>0$ such that $\lambda_l<M$, for all $l$, then
$P_n(\mathcal{C})\rightarrow0$ as $n\rightarrow\infty$.

Let $\psi_n(k)$ denote the probability that $G(r,\Lambda)$ consists
of $k$ components and $P_n^m(k)$ the probability that there are $k$
components in $G(r,\Lambda)$, each of which having size $m$ (i.e.
$m$ vertices).

\textit{Proposition 4:} Suppose there exists $M>0$ such that
$\lambda_l<M$, for all $l$. Then, for any fixed $k$,
$\psi_n(k)\rightarrow0$ as $n\rightarrow\infty$; and for any fixed
$k$, $m$, $P_n^m(k)\rightarrow0$ as $n\rightarrow\infty$.

\textit{Proof:} Mimicking the proof of Theorem 3 and 4 in \cite{10}
yields the result. \hspace*{\fill}~\QED

\begin{figure}[!t]
\centering \scalebox{0.5}{\includegraphics{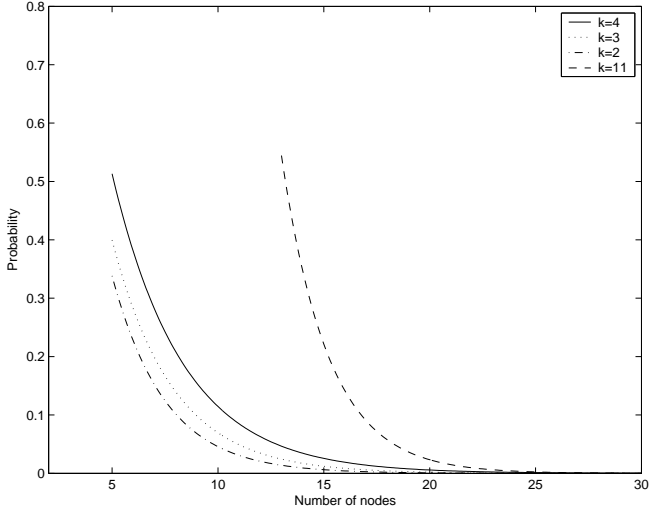}}
\caption{Probability that $G(r,\Lambda)$ contains $k$ components for
different values of $k$} \label{fig_sim}
\end{figure}

In Figure 2, we plot $\psi_n(k)$ as a function of $n$ number of
vertices for different $k$. We take $\lambda_i=1$ for $1\le i\le
10$, and $\lambda_i=2$ for $i>10$. Observe that the convergence to
the asymptotic value 0 is very fast.

We may thus conclude that this static network is almost surely
divided into an infinite number of finite clusters. This observation
was first made in \cite{2} by a different approach.

\subsection{Degree Distribution}

Let $G(r,\lambda)$ denote the graph $G(r,\Lambda)$ when
$\Lambda=\{\lambda,\cdots,\lambda\}$.

\textit{Theorem 4:} In the graph $G(r,\lambda)$, the degree
distribution can be divided into three classes: the degree
distribution of $X_1$ and $X_n$ is $Poi(\lambda r)$; and for $k+1\le
i\le n-k$, that of $X_i$ is $\big\{e^{-2\lambda r}\frac{(2\lambda
r)^k}{k!}\big\}_{k\in\mathbb{N}}$. For $2\le i\le k$, the degree
distribution of $X_i$ and $X_{n+1-i}$ is $\big\{e^{-2\lambda
r}\frac{(\lambda r)^k}{k!}\sum_{j=0}^{i-1}{k \choose
j}\big\}_{k\in\mathbb{N}}$.

\textit{Proof:} Let $\{Y_i\}$, $\{Y'_i\}$ be independent
$Exp(\lambda)$. Denote the degree of vertex $X_i$ as $d_i$. We get
\begin{eqnarray*}
P(d_n\ge k)&=&P(d_1\ge k)=P(Y_1+\cdots+Y_k\le r)\\
&=&e^{-\lambda r}\Big(\frac{(\lambda r)^k}{k!}+\frac{(\lambda
r)^{k+1}}{(k+1)!}+\cdots\Big),
\end{eqnarray*}
where we used an equivalent definition of gamma distribution. Hence,
$$
P(d_n=k)=P(d_1=k)=e^{-\lambda r}\frac{(\lambda r)^k}{k!}.
$$
Next, for $2\le i\le k$,
\begin{eqnarray*}
P(d_{n+1-i}=k)&=&P(d_i=k)\\
&=&\sum_{j=0}^{i-1}P(Y_1+\cdots+Y_j\le
r,\\
& &Y_1+\cdots+Y_{j+1}>r)\\
  & &\cdot P(Y'_1+\cdots+Y'_{k-j}\le
r,\\
& &Y'_1+\cdots+Y'_{k-j+1}>r)\\
&=&\sum_{j=0}^{i-1}\int_0^r\lambda e^{-\lambda x}\frac{(\lambda
x)^{j-1}}{(j-1)!}\\
& &\int_{r-x}^{\infty}\lambda e^{-\lambda
y}\mathrm{d}y\mathrm{d}x\\
& &\cdot\int_0^r\lambda e^{-\lambda x}\frac{(\lambda
x)^{k-j-1}}{(k-j-1)!}\\
& &\int_{r-x}^{\infty}\lambda
e^{-\lambda y}\mathrm{d}y\mathrm{d}x\\
&=&e^{-2\lambda r}\frac{(\lambda r)^k}{k!}\sum_{j=0}^{i-1}{k \choose
j}.
\end{eqnarray*}
Finally, for $k+1\le i\le n-k$,
\begin{eqnarray*}
P(d_i=k)&=&\sum_{j=0}^kP(Y_1+\cdots+Y_j\le
r,\\
& &Y_1+\cdots+Y_{j+1}>r)\\
& &\cdot P(Y'_1+\cdots+Y'_{k-j}\le
r,\\
& &Y'_1+\cdots+Y'_{k-j+1}>r)\\
&=&e^{-2\lambda r}\frac{(2\lambda r)^k}{k!}
\end{eqnarray*}
which concludes the proof. \hspace*{\fill}~\QED

\subsection{Strong Law Results}

Define the connectivity distance $c_n:=\inf\{r>0:G(r,\lambda)\
\mathrm{is}\ \mathrm{connected}\}$; and the largest nearest neighbor
distance $b_n:=\max_{1\le i\le n}\min_{1\le j\le
n,j\not=i}\{|X_i-X_j|\}$. We derive asymptotic tight bounds for
$c_n$ and strong law of large numbers for $b_n$, as $n$ tends to
infinity.

\textit{Theorem 5:}
In the graph $G(r,\lambda)$, we have\\
(i)
$$
\limsup_{n\rightarrow\infty}\frac{\lambda c_n}{\ 2\ln n}\le 1\quad
\mathrm{and}\quad\liminf_{n\rightarrow\infty}\frac{\lambda c_n}{\
\ln n}\ge 1\qquad a.s.
$$
(ii)
$$
\lim_{n\rightarrow\infty}\frac{\lambda b_n}{\ \ln n}=1\qquad a.s.
$$

\textit{Proof:} (i) Observe that $P(c_n\ge
x)\le\sum_{l=1}^{n-1}e^{-\lambda_lx}=(n-1)e^{-\lambda x}$ invoking
the Boole inequality. Let $\varepsilon>0$. Take
$x=x_n=(2+\varepsilon)\ln n/\lambda$ in the above expression and sum
in $n$, then we get
$$
\sum_{n=1}^{\infty}P(c_n\ge x_n)\le
\sum_{n=1}^{\infty}n^{-(1+\varepsilon)}<\infty.
$$
By the Borel-Cantelli lemma, $P(c_n\ge x\ \mathrm{i.o.})=0$. Hence,
$\limsup_{n\rightarrow\infty}\frac{\lambda c_n}{\ 2\ln n}\le 1$
almost surely.

On the other hand, $P(c_n\le y)=\prod_{l=1}^{n-1}(1-e^{-\lambda_l
y})=(1-e^{-\lambda y})^{n-1}$. Take $y=y_n=(1-\varepsilon)\ln
n/\lambda$, then
$$
\sum_{n=1}^{\infty}P(c_n\le
y_n)\le\sum_{n=1}^{\infty}\big(1-n^{-(1-\varepsilon)}\big)^{n-1}\sim
\sum_{n=1}^{\infty}e^{-n^{\varepsilon}}<\infty.
$$
We conclude that $\liminf_{n\rightarrow\infty}\frac{\lambda c_n}{\
\ln n}\ge 1$ a.s. by using the Borel-Cantelli lemma again.

(ii) By the independence of $\{Y_l\}$, we obtain
\begin{eqnarray*}
P(b_n\ge x)&=&P\big(\cup_{i=2}^{n-1}\{\{Y_{i-1}\ge x\}\cap\{Y_i\ge
x\}\}\\
& &\cup\{Y_1\ge x\}\cup\{Y_{n-1}\ge
x\}\big)\\
&\le&\sum_{i=2}^{n-1}P(Y_{i-1}\ge x)\cdot P(Y_i\ge x)\\
& &+P(Y_1\ge
x)+P(Y_{n-1}\ge x)\\
&=&(n-2)e^{-2\lambda x}+2e^{-\lambda x}.
\end{eqnarray*}
Take $x=x_n=(2+\varepsilon)\ln n/(2\lambda)$, then we get
$$
\sum_{n=1}^{\infty}P(b_n\ge
x_n)\le\sum_{n=1}^{\infty}\big(n^{-(1+\varepsilon)}+2n^{-(1+\frac{\varepsilon}2)}\big)<\infty.
$$
By the Borel-Cantelli lemma,
$\limsup_{n\rightarrow\infty}\frac{\lambda b_n}{\ \ln n}\le 1$
almost surely.

On the other hand,
\begin{eqnarray*}
P(b_n\le y)&=&P\big(\cap_{i=2}^{n-1}\{\{Y_{i-1}\le y\}\cup\{Y_i\le
y\}\}\\
& &\cap\{Y_1\le y\}\cap\{Y_{n-1}\le
y\}\big)\\
&\le&\prod_{i=1}^{\lfloor\frac n2\rfloor}P(Y_{2i-1}\le y)\cdot
P(Y_{2i}\le y)\\
&\sim&(1-e^{-\lambda y})^n.
\end{eqnarray*}
Arguing similarly as in (i), we can get
$\liminf_{n\rightarrow\infty}\frac{\lambda b_n}{\ \ln n}\ge 1$ a.s..
This completes the proof. \hspace*{\fill}~\QED

\section{Concluding Remarks}

This paper dealt with random geometric graphs in one dimension in
which the vertex positions were evolving time. The critical assumption
that this evolution was modeled by describing an evolution
equation for the change in the inter-nodal spacing. We studied some
dynamical as well as static properties and results were given for fixed
$n$ total number of vertices as well as $n$ tending to infinity.

It is worth pointing out that this paper is only a preliminary step
on the investigation of exponential RGG process models. The idea of
considering spacings may be extended to high dimensions in the
following way. Deploy $X_1$ according to a probability density $f$,
then place $X_2$ with the same probability density substituting the
location of $X_1$ for the coordinate origin, and so forth. We deem that
the growing scheme would be an important alternative from the
typical binomial and Poisson cases \cite{15}. Other interesting
directions include examination of ``multiple spacings'', reinforcing
1-step memory to finite steps memory even to infinite one, which could
be possible to result in power law degree distributions. Since we
only treat the limit regime for constant $\lambda_l$, how to deal
with $\lambda_l$ approaching infinity is our future research.

\end{document}